\documentclass[11pt,twoside]{article}

\usepackage{asp2006}
\usepackage{graphicx}
\usepackage{lscape}

\begin{document}

\title{Mass Loss History of the AGB star, R Cas}   
\author{%
Toshiya Ueta$^1$,
Robert E.\ Stencel$^1$, 
Issei Yamamura$^2$, 
Hideyuki Izumiura$^3$,
Yoshikazu Nakada$^{4,5}$,
Mikako    Matsuura$^{6}$,
Yoshifusa Ita$^{2,6}$,
Toshihiko Tanab\'{e}$^{4}$,
Hinako    Fukushi$^{4}$,
Noriyuki  Matsunaga$^{6}$,
Hiroyuki  Mito$^{5}$, and
Angela K.\ Speck$^7$}   

\affil{%
$^1$ Dept.\ of Physics and Astronomy, University of Denver, USA \\
$^2$ Institute of Space and Astronautical Science, JAXA \\
$^3$ Okayama Astrophysical Observatory, NAOJ\\
$^4$ Institute of Astronomy, Schooof Science, University of Tokyo\\
$^5$ Kiso Observatory, Institute of Astronomy, University of Tokyo\\
$^6$ National Astronomical Observatory of Japan\\
$^7$ Dept.\ of Physics \& Astronomy, University of Missouri, USA}

\begin{abstract} 
We report here on the discovery of an extended far-infrared shell around
 the AGB star, R Cassiopeia, made by {\sl AKARI} and {\sl Spitzer}.
The extended, cold circumstellar shell of R Cas spans nearly $3\arcmin$
 and is probably shaped by interaction with the interstellar medium.  
This report is one of several studies of well-resolved mass loss
 histories of AGB stars under {\sl AKARI} and {\sl Spitzer} observing
 programs labeled ``Excavating Mass Loss History in Extended Dust
 Shells of Evolved Stars (MLHES)''.  
\end{abstract}


\section{Introduction}   
\citet{deutsch56} discussed the existence of
blueshifted circumstellar cores in the spectrum of the red supergiant
star $\alpha$ Her, constituting direct evidence for high rates of
mass loss.
Since that time, numerous observations have elucidated the magnitude and
ubiquity of mass loss across the upper right side of the
Hertzsprung-Russell diagram.  
The high rates of mass loss among AGB stars rival evolutionary
timescales, affecting stellar evolutionary tracks substantially. 
Therefore, the high rates of mass loss deserve careful determination.  
These facts have befuddled theorists, who are struggling with the basic
challenge of how to lift so much mass away from the gravitational hold
of the star. 

During the 1980's, observations with the {\sl Infrared Astronomical
Satellite} ({\sl IRAS}\/) demonstrated that extended infrared shells of
evolved stars - the anticipated effect of continuous dusty mass loss
(e.g.\ \citealt{Sten88,Y93a,Y93b}) -  were present.  
During the 1990's, the {\sl Infrared Space Observatory} 
and ground-based infrared work began to refine those results
(e.g.\ \citealt{Izu97,Meixner99}), indicating variations in the mass loss 
rate over time and resulting in shells and structure as might be
predicted in stellar evolution calculations for AGB stars (e.g.\
\citealt{Iben95}).    
This decade, we are fortunate to have higher resolution and sensitivity
instruments like the {\it AKARI Infrared Astronomy Satellite} ({\sl
AKARI}: \citealt{Murakami07}) and the {\it Spitzer Space Telescope}
({\it Spitzer}: \citealt{Wer04}) that can more carefully map the mass
loss history of evolved stars.  

This report is the first of several studies of well-resolved mass loss
histories of AGB stars under {\it AKARI} and {\it Spitzer} observing
programs labeled ``Excavating Mass Loss History in Extended Dust Shells
of Evolved Stars (MLHES)''. 
In this contribution, we will explore whether the detection of an extended
($\sim 3^{\prime}$ radius) dust shell around the oxygen-rich AGB star, R
Cas, can be interpreted in terms of thermal pulses, in order to
constrain mass loss histories from this evolved star.  
In parallel with the mass loss history of evolved stars, evidence for
the interaction of circumstellar and interstellar medium (ISM) is growing,
with new observations of bow shocks around R Hya \citep{ueta06} and Mira
\citep{Martin07,Ueta08}, plus theoretical considerations of the
phenomena \citep{Villaver03,wareing07}.  
Because nature is intrinsically complicated, we will examine evidence
for both mass loss history and interaction with the ISM in the case
of the extended infrared shell of R Cas.  

\section{R Cas: the Star and its Circumstellar Shell} 
The Mira type variable, R Cas (HD224490), is an oxygen-rich star with a
period of 431 days and estimated mass loss rate of 10$^{-7}$
M$_{\odot}$ yr$^{-1}$.  
This star shows an extended circumstellar shell originally detected by
{\sl IRAS} at $60\micron$, having angular extent at least $4\farcm3$
\citep{Y93a}, or $2\farcm8$ \citep{BS94} using careful 
deconvolution of the point-spread-function (PSF). The linear extent of
the shell depends on the distance determination.   

We observed R Cas in the four bands at 65, 90, 140 and $160\micron$
using the Far-Infrared Surveyor (FIS: \citealt{kawada07}) on-board 
{\sl AKARI\/} on 2007
January 16 as part of the MLHES Mission Program (PI: I.\ Yamamura). 
Observations were made with the FIS01 (compact source photometry) scan
mode, in which two strips of forward and backward scans (at $0.5$s reset
rate) were done with a 70$^{\prime\prime}$ spacing.

R Cas was also observed at $70\micron$ using MIPS aboard {\sl Spitzer} on
2008 February 18 as part of the Cycle 4 GO project, {\sl Spitzer}-MLHES
(PID 40092, PI: T.\ Ueta). 
We mapped a $15^{\prime} \times 20^{\prime}$ region with a series of
exposures in photometry/fixed-cluster-offset mode, while avoiding the 
central star that is brighter than the saturation limit of the MIPS
arrays.

\begin{figure}[!ht]
\begin{center}
   \resizebox{1.0\hsize}{!}{
     \includegraphics*{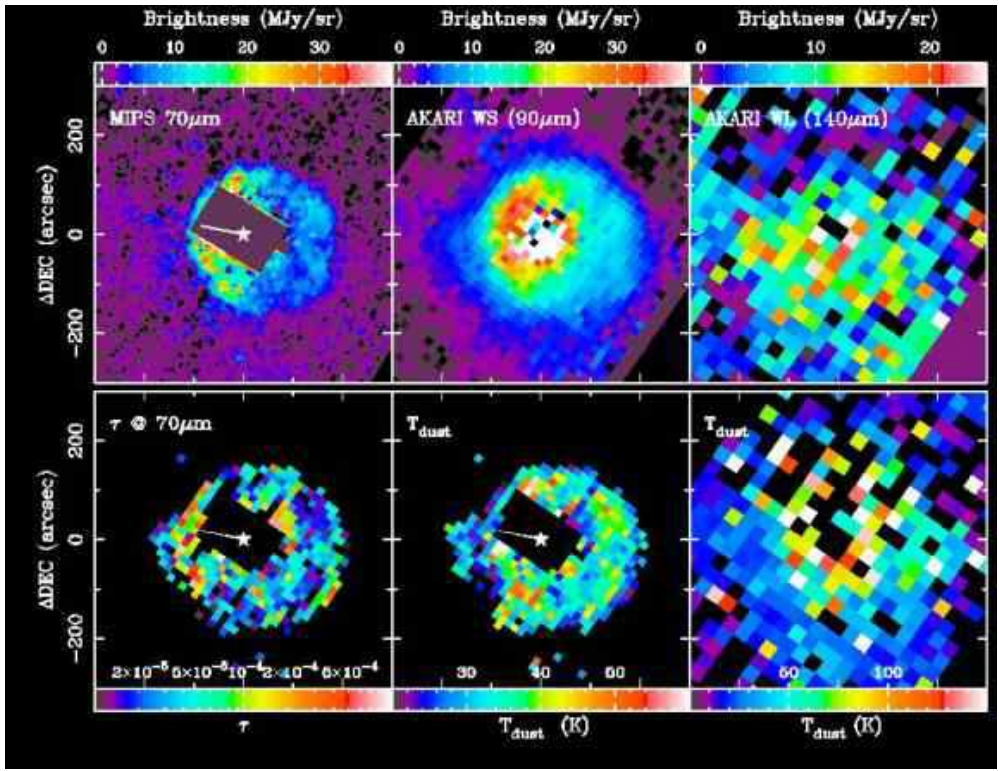}
   }
\end{center}
\caption{R Cas images in the far-infrared and derived quantities. [Top]
 {\sl Spitzer} $70\micron$ (PSF subtracted, left) and {\sl AKARI}
 $90\micron$ (WIDE-S, middle) and $140\micron$ (WIDE-L, right) maps.
 [Bottom] Optical depth map at $70\micron$ (left) and dust temperature
 maps derived from {\sl Spitzer} and AKARI $90\micron$ map (middle) and
 from AKARI $90$ and $140\micron$ maps (right). The line emanating from
 the central star's position indicates the direction of the star's
 proper motion.
The color image is available from the version on arXiv.}  
\label{fig:RCas}
\end{figure}

\section {Preliminary Results and Discussion} 

Figure \ref{fig:RCas} shows a {\sl Spitzer} image at $70\micron$
(top left) and {\sl AKARI} images at 90 and $140\micron$ (top middle and
right, respectively). 
These far-infrared images show the extended circumstellar shell around R
Cas, which is roughly circular of radius 140 to $165\arcsec$.
The shell's emission structure consists of the relatively flat
``plateau'' region on the west side (of surface brightness $\sim 15$ to
20 MJy sr$^{-1}$ at 65 to 90$\mu$m and $< 10$ MJy sr$^{-1}$ at
140$\mu$m) and the region of higher surface brightness (emission core)
on the east side around the central star. 
We immediately see that the extended shell is off-centered, i.e.,
position of the central star does not coincide with the geometric center
of the extended circular shell.
Also, there appears to be an enhancement of the surface brightness along
the periphery of the shell (especially recognizable in the {\sl
Spitzer} $70\micron$ and the dust temperature maps).
This density enhancement may be due to a fast low density wind
shock-merging with a slower, high density wind which was caused during
an era of higher/enhanced mass loss, as has been hydrodynamically
predicted (e.g.\ \citealt{Steffen98}). 

{\sl Hipparcos} measured the proper motion of $(\mu_{\alpha},
\mu_{\delta})  = (85.5\pm 0.8, 17.5\pm 0.7)$ mas yr$^{-1}$
\citep{leeuwen07}. 
This translates to the position angle of $78\fdg4 \pm 0\fdg7$ east of
north, which is indicated as a line drawn from the position of the star
in Figure \ref{fig:RCas}.
This direction agrees remarkably well with the direction along which
there is a positive gradient of surface brightness.
In other words, the brighter surface brightness in the eastern edge of
the shell appears to arise from interactions between AGB winds and the
ISM, where a bow shock is expected as discovered around another AGB
star, R Hya \citep{ueta06}.  

The shape of the circumstellar shell was fit by an ellipse.
The best-fit semi-major axis and eccentricity pair, $(a,\epsilon)$,  is
$a = 33.6$ pix $= 165\farcs3$ and $\epsilon = 0.3$: indeed, the R Cas
shell is not quite circular in projection. 
According to the best-fit ellipse, the distance from the ellipse center
to one of the foci is $c = a \times \epsilon = 9.9$ pix $= 48\farcs8$.
This means that the central star is $48\farcs8$ displaced from the
ellipse center over the course of the shell expansion.  
At the preferred distance for R Cas, 176 pc, the semi-major and
semi-minor axes correspond to 0.12 to 0.14 pc.
At the measured CO expansion rate of 12 km s$^{-1}$ \citep{bfo94}, a crossing
time of the shell is therefore roughly $10^4$ years.  

If the elongation of the R Cas shell is solely due to the motion of the
central star with respect to the shell (in an otherwise stationary local
environment), the star must have been moving roughly at 5 mas yr$^{-1}$, which
is much smaller than the observed motion of 86.5 mas yr$^{-1}$.
Thus, the shaping of the shell is NOT self-inflicted as in interactions
between fast and slow AGB winds expected in the AGB shells
\citep{Steffen98}. 
Rather, the shall shaping appears to be a result of interactions between
AGB winds emanating from the moving star AND the ISM flow local to R
Cas.
As shown by \citet{ueta09} in this volume, one can deduce the direction
of the ISM flow local to R Cas given the direction of the proper motion
by fitting the shell/bow structure in the leading, brighter eastern
edge.
While similar analyses would allow probing of the 3-D dynamics of the
ISM local to these wind-ISM interacting shells, these shells do not
preserve mass loss history beyond the wind crossing time of the shell.
In the case of R Cas, mass loss history can be probed only up to $10^4$
yr ago from the internal structure of the shell.

\acknowledgements 
This research is based on observations with {\sl AKARI}, a JAXA
project with the participation of ESA, and {\sl Spitzer}, which is
operated by the JPL/Caltech under a contract with NASA.


\end{document}